\documentclass[]{aa} 
\begin{document}
\outer\def\gtae {$\buildrel {\lower3pt\hbox{$>$}} \over 
{\lower2pt\hbox{$\sim$}} $}
\outer\def\ltae {$\buildrel {\lower3pt\hbox{$<$}} \over 
{\lower2pt\hbox{$\sim$}} $}
\newcommand{\ergscm} {ergs s$^{-1}$ cm$^{-2}$}
\newcommand{\ergss} {ergs s$^{-1}$}
\newcommand{\ergsd} {ergs s$^{-1}$ $d^{2}_{100}$}
\newcommand{\pcmsq} {cm$^{-2}$}
\newcommand{\ros} {\sl ROSAT}
\newcommand{\exo} {\sl EXOSAT}
\newcommand{\xmm} {\sl XMM-Newton}
\newcommand{\chan} {\sl Chandra}
\def\rchi{{${\chi}_{\nu}^{2}$}}
\def\uchi{{${\chi}^{2}$}}
\newcommand{\Msun} {$M_{\odot}$}
\newcommand{\Mwd} {$M_{wd}$}
\def\Mdot{\hbox{$\dot M$}}
\def\mdot{\hbox{$\dot m$}}
\input psfig.sty
%
\title{\sl{XMM-Newton} observations of AM CVn binaries: V396 Hya \& 
SDSS J1240--01}
\titlerunning{\sl{XMM-Newton} observations of AM CVn binaries: V396 Hya \& 
SDSS J1240--01}
\authorrunning{Ramsay et al} 


\author{Gavin Ramsay \inst{1}, Paul J. Groot\inst{2}, Tom Marsh$^{3}$, 
Gijs Nelemans$^{2}$, Danny Steeghs$^{4}$, Pasi Hakala\inst{5,6}}

\offprints{G. Ramsay: gtbr@mssl.ucl.ac.uk}

\institute{
$^{1}$Mullard Space Science Laboratory, University College London,
Holmbury St Mary, Dorking, Surrey, RH5 6NT, UK.\\
$^{2}$Department of Astrophysics, IMAPP, Radboud University of Nijmegen, PO
Box 9010, 6500 Nijmegen, The Netherlands\\
$^{3}$Department of Physics, University of Warwick, Coventry, CV4 7AL, UK\\ 
$^{4}$Harvard-Smithsonian Center for Astrophysics, 60 Garden
Street, MS-67, Cambridge, MA 02138, USA\\ 
$^{5}$Observatory, P.O. Box 14, FIN-00014 University of Helsinki,
Finland.\\
$^{6}$Tuorla Observatory, V\"ais\"al\"antie 20, 21500 Piikki\"o, Finland\\
}


\date{Accepted: 1 July 2006}

\abstract{ We present the results of {\xmm} observations of two AM CVn
systems -- V396 Hya and SDSS J1240-01. Both systems are detected in
X-rays and in the UV: neither shows coherent variability in their
light curves. We compare the rms variability of the X-ray and UV power
spectra of these sources with other AM CVn systems. Apart from ES Cet,
AM CVn sources are not strongly variable in X-rays, while in the UV
the degree of variability is related to the systems apparent
brightness. The X-ray spectra of V396 Hya and SDSS J1240-01 show
highly non-solar abundances, requiring enhanced nitrogen to obtain
good fits. We compare the UV and X-ray luminosities for 7 AM CVn
systems using recent distances. We find that the X-ray luminosity is
not strongly dependent upon orbital period. However, the UV luminosity
is highly correlated with orbital period with the UV luminosity
decreasing with increasing orbital period. We expect that this is due
to the accretion disk making an increasingly strong contribution to
the UV emission at shorter periods. The implied luminosities are in
remarkably good agreement with predictions.  }

\maketitle

\section{Introduction}

Ultra-compact AM CVn binaries consist of a white dwarf which accretes
from a degenerate (or semi-degenerate) companion star in a very short
period binary. They are observed to have orbital periods shorter than
$\sim$70 mins. These systems allow us to observe accretion flows which
are hydrogen-deficient, which provides an important comparison with
their hydrogen-dominated counterparts (the `classical' cataclysmic
variables).

Until very recently they were not well studied in X-rays or in the UV.
Ramsay et al (2005) presented {\xmm} observations of 4 AM CVn system
which showed that each system was easily detected at X-ray
energies. Further, they showed that a large proportion of the
accretion energy of AM CVn systems was emitted in the UV. Observations
of the shortest period AM CVn, ES Cet, shows that it too is seen in
X-rays (Strohmayer 2004b). (The two systems with shorter periods, RX
J0806+15 at 5.3 min and RX J1914+24 at 9.5 min remain candidate AM CVn
systems).

Until now, the sample of AM CVn systems which have been observed using
dedicated X-rays observations has been biased towards shorter orbital
periods. To rectify this bias we have obtained observations of two AM
CVn systems which have relatively long orbital periods. V396 Hya (also
known as CE 315, Ruiz et al 2001a), with a period of 65.2 min has the
longest orbital period of known AM CVn systems, while SDSS
J124058.03-015919.2 (hereafter SDSS J1240-01, Roelofs et al, 2004,
2005) has the third longest (37.4 min).

\section{Observations}

{\xmm} has 3 broad-band X-ray detectors with medium energy resolution
and also a 30 cm optical/UV telescope (the Optical Monitor, OM: Mason
et al 2001). The X-ray instruments contain imaging detectors covering
the energy range 0.15--10keV; one EPIC pn detector (Str\"{u}der et al
2001) and two EPIC MOS detectors (Turner et al 2001).  Two high
resolution grating spectrometers (the RGS) are also on board, but our
sources are too faint to obtain useful spectra. The observation log is
shown in Table \ref{log} where we show the mean X-ray and UV count
rates for each source.

The data were processed using the {\sl XMM-Newton} {\sl Science
Analysis Software} (SAS) v6.5 and analysed in a similar manner to that
described in Ramsay et al (2005). For the OM observations, we used the
UVW1 filter which has a central wavelength of 2910\AA\hspace{1mm} and
a range of 2400--3400\AA. In the case of V396 Hya, the source was just
outside the fast window mode which has a small field of view (V396 Hya
also has a significant proper motion, Ruiz et al 2001b). It was,
however, seen in full window mode which gives the mean brightness over
a time interval of typically 4400 sec. In contrast SDSS J1240-01 was
observed in the OM fast window mode.

\begin{table}
\begin{center}
\begin{tabular}{lcrrr}
\hline
Source & Date & Exp & EPIC  & UV \\
       &      & (ksec) & Ct/s & Ct/s \\
\hline
V396 Hya & 2005-07-20 & 27.9 & 0.38 & 1.65 \\
SDSS J1240-01 & 2006-01-07 & 22.2 & 0.023 & 0.44 \\
\hline
\end{tabular}
\end{center}
\caption{The observation log for V396 Hya and SDSS J1240-01. The exposure 
time is that of the EPIC pn detector. The count
rate refers to the mean count rate in the EPIC pn detector
(0.15--10keV) and the UVW1 filter (2400-3400\AA).}
\label{log}
\end{table}

\section{Light curves}

V396 Hya was relatively bright in X-rays (Table \ref{log}) giving an
apparent count rate greater than that of AM CVn and CR Boo, but less
than HP Lib and GP Com (Ramsay et al 2005). In contrast SDSS J1240-01
is fainter than all of the above, and only slightly brighter than ES
Cet (Steeghs et al 2006). There is evidence for some structure in the
X-ray light curve of V396 Hya, but little in SDSS J1240-01. In the UV,
both sources were relatively bright, but fainter than the AM CVn
sources reported by Ramsay et al (2005) and also ES Cet (Steeghs et al
2006). We searched for periodic variations in the X-ray and UV light
curves of V396 Hya and SDSS J1240-01, but found none.

\begin{figure}
\begin{center}
\setlength{\unitlength}{1cm}
\begin{picture}(8,5.8)
\put(-0.2,-0.2){\includegraphics{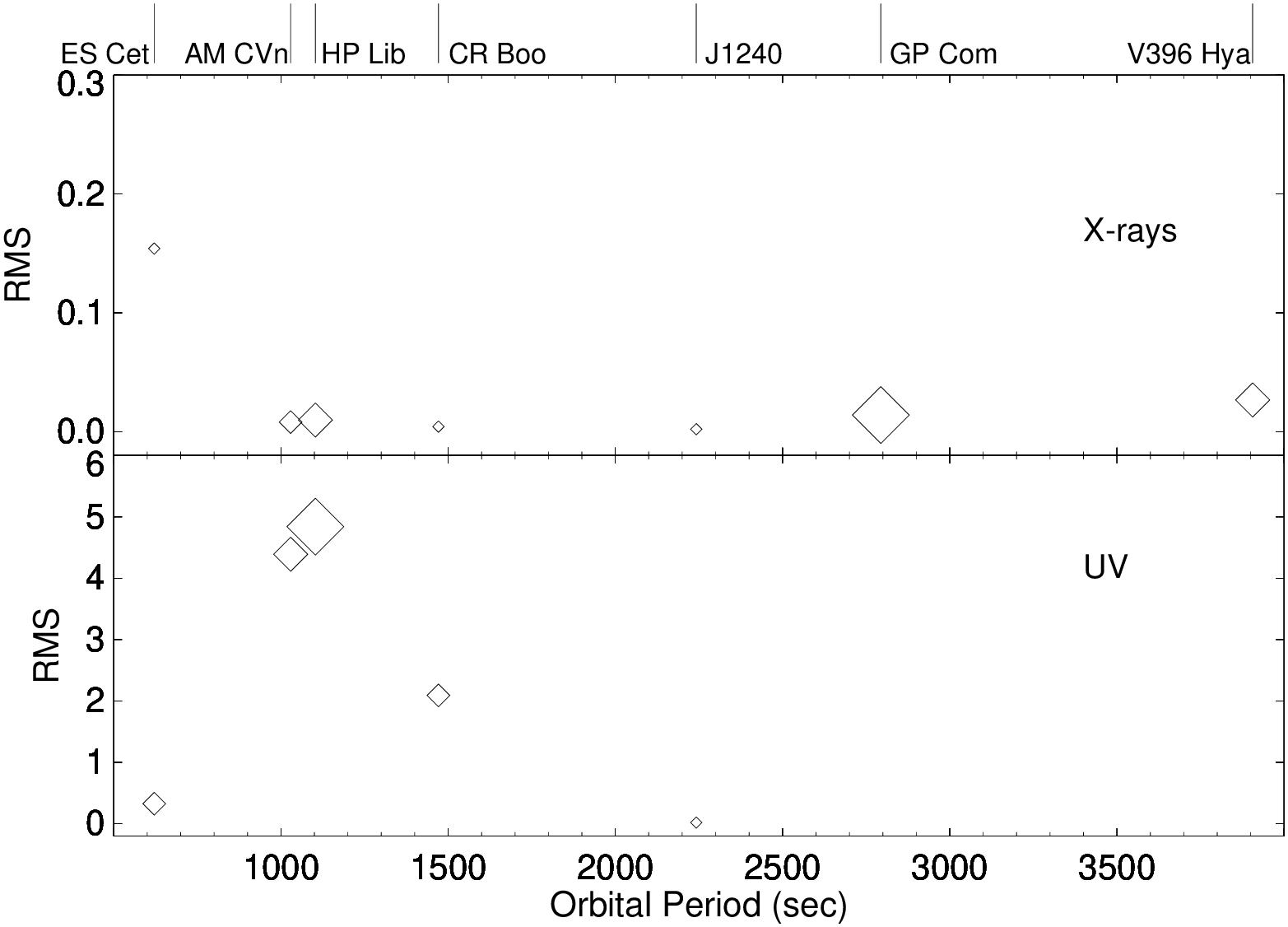}}
\end{picture}
\end{center}
\caption{The rms of the amplitude spectra of the light curves of AM
CVn systems in the 0.1-1.0keV X-ray band (top panel) and the UV
(bottom panel). The size of the symbol reflects the apparent
brightness in that energy. No UV fast mode data were obtained of GP
Com and V396 Hya.}
\label{variability}
\end{figure}

To determine the level of variability for our wider sample of AM CVn
systems, we measured the rms of the power spectra for each of the
X-ray and UV light curves.  We did this for the 0.15-1keV energy band
and the UVW1 filter using light curves with 10 sec time bins. The
results are shown in Figure \ref{variability} where the size of the
symbol reflects the relative brightness in X-rays and the UV.  In the
X-ray band, there is no evidence for a correlation between the degree
of variability and orbital period apart from ES Cet, which has the
shortest orbital period and shows the largest rms. In the UV, the
degree of variability appears to be related to the apparent brightness
of the system, with the rms increasing with increasing brightness.

Our variability measure reflects continuum variability at only one
epoch. For instance GP Com showed large variations in its UV line
emission which was taken to be due to irradiation of the accretion
disk or variable mass loss in a wind (Marsh et al 1995). Further, van
Teeseling \& Verbunt (1994) showed a significant variation in the soft
X-ray light curve of GP Com in {\ros} observations.

\section{Spectra}

The X-ray spectra of the AM CVn systems in our earlier sample were
best fitted using a relatively low temperature thermal plasma model
with non-solar metallicities (eg Strohmayer 2004a, Ramsay et al
2005). UV Observations of V396 Hya made using the {\sl HST} also show
a significant enhancement of nitrogen (G\"{a}nsicke et al 2003).  We
extracted spectra in the same manner as before, but filtered out time
intervals of enhanced background.  We fitted the spectra using the
{\tt cevmkl} model in {\tt XSPEC} together with a neutral absorption
model. The fits to the spectra using abundances expected from
relatively low temperature CNO processed material were poor: for the
EPIC pn spectra the fits were \rchi=3.52 (173 dof) for V396 Hya and
\rchi=2.44 (12 dof) for SDSS J1240-01.

\begin{figure*}
\begin{center}
\setlength{\unitlength}{1cm}
\begin{picture}(8,5)
\put(-5,-0.9){\includegraphics{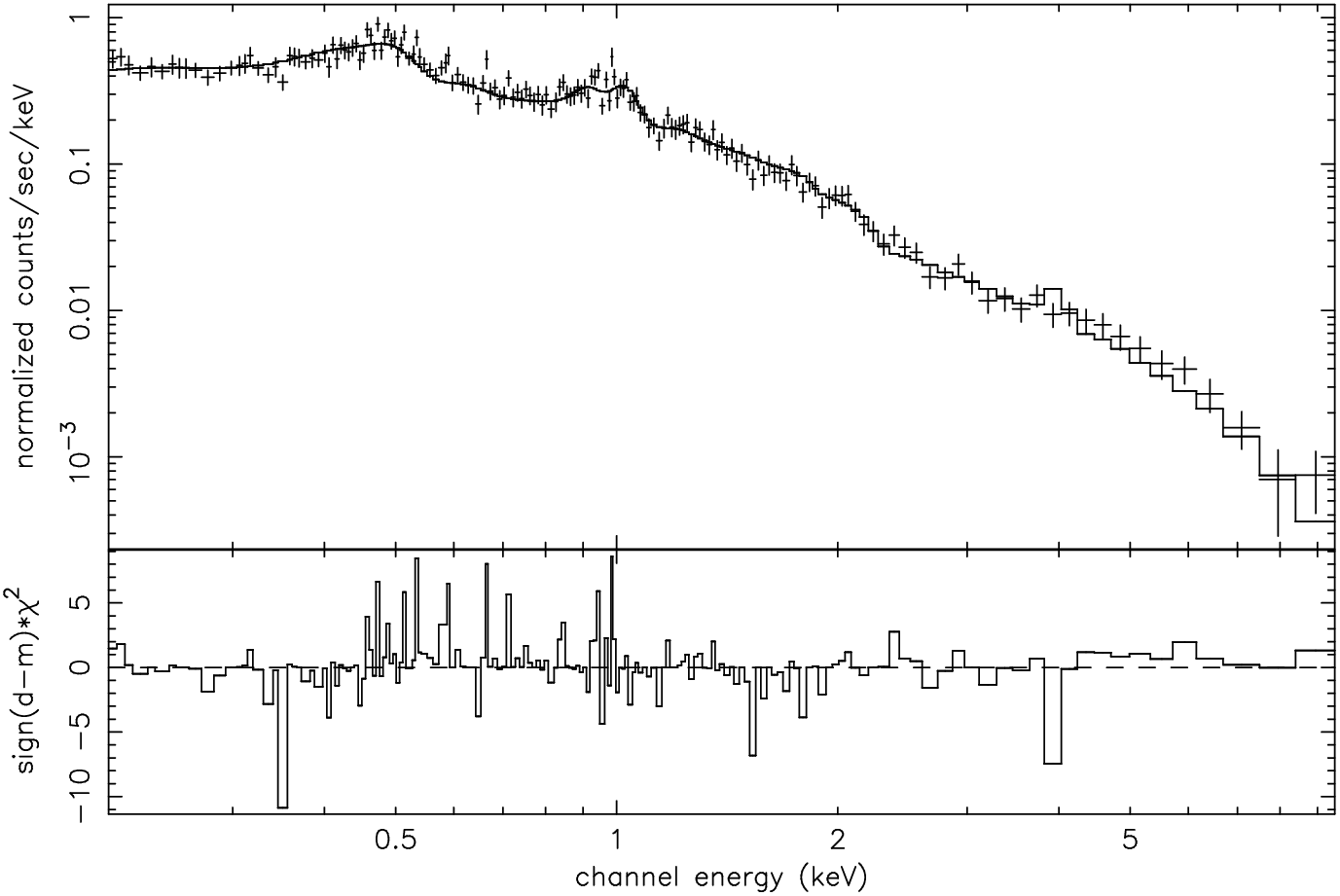}}
\put(4,-0.9){\includegraphics{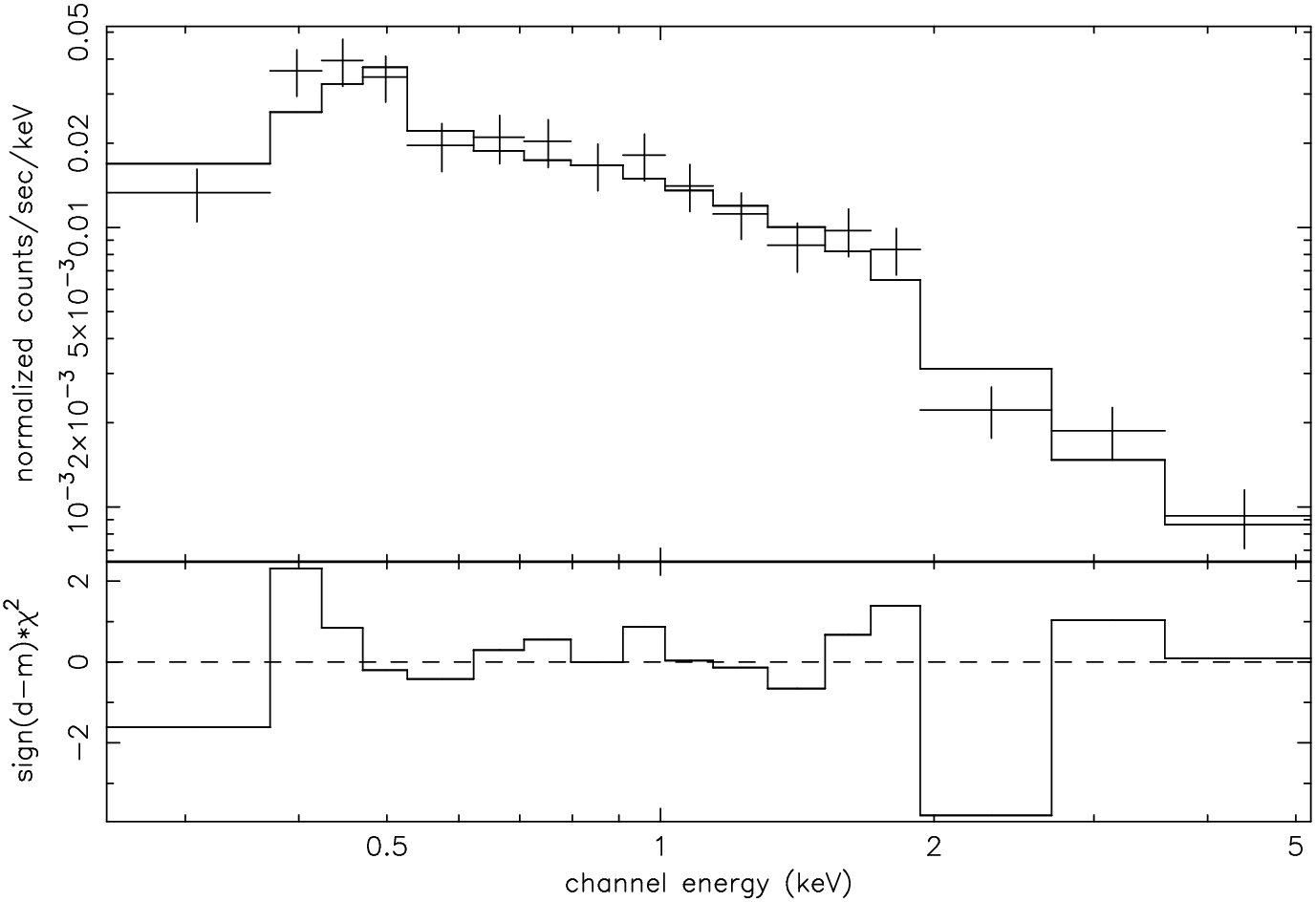}}
\end{picture}
\end{center}
\caption{The EPIC pn spectra for V396 Hya (left panel) and SDSS
J1240-01 (right panel) together with the best model fits using an
absorbed thermal plasma model with variable metal abundances.}
\label{spec}
\end{figure*}

We then set the abundances of each element to zero (ie a completely
ionised plasma) and then allowed each element to vary. If there was no
significant improvement to the fit, then the abundance of that element
was set back to zero and frozen. For both V396 Hya and SDSS J1240-01
we find that a significant amount of N improved the fits, while in the
case of V396 Hya a significant amount of Ne and less significant
amount of Ca also improved the fits. Fits using the EPIC MOS data were
in agreement with these findings. The EPIC pn spectra of each source
together with their best fits are shown in Figure \ref{spec} while the
best fit spectral parameters are shown in Table \ref{bestfits}.

To test the robustness of these results we also fitted the X-ray
spectra assuming solar abundances apart from hydrogen which was set to
zero. Each element was then allowed to vary in abundance: if this
resulted in no significant difference to the fit, it was re-fixed to
the solar value. In the case of SDSS J1240-01 we find that nitrogen
with an abundance of 55\% greater than solar gave a better fit to the
spectrum (at the 99.1\% confidence level) and in the case of V396 Hya
at an amount of 68-92\% greater than solar (at $>$99.99\%
confidence). Other spectral parameters, such as the amount of
absorption and the maximum temperature of the plasma, were similar to
that found before (cf Table \ref{bestfits}). In this test, we found no
evidence for an enhancement of neon in the X-ray spectrum of V396
Hya. We conclude that nitrogen is significantly enhanced in the X-ray
spectra of V396 Hya and SDSS J1240-01.

\begin{table*}
\begin{center}
\begin{tabular}{lrrrrrrrr}
\hline
Source & $N_{H} $ & $T_{max}$ & Z & $F_{x,o}$ &
$F_{x,u}$ & $L_{X}$ & \rchi (dof)\\
       & $\times10^{20}$ & (keV) & (solar) & \ergss & \ergss & erg/s \\
       & \pcmsq &  &  & cm$^{-2}$ & cm$^{-2}$ &  \\
\hline
V396 Hya & 3.4$^{+0.9}_{-0.6}$ & 5.5$_{-0.7}^{+0.5}$ & 
6$\pm1.2$ (N) & 1.05$^{+0.11}_{-0.07}\times10^{-12}$ & 
2.42$^{+0.24}_{-0.16}\times10^{-12}$ & 
1.7$^{+0.2}_{-0.1}\times10^{30}$ & 1.30 (170)\\
  & & & 0.8$\pm0.2$ (Ne) & & & & \\
SDSS J1240-01 & 9.0$^{+7}_{-0.5}$ & 6.1$^{+8.3}_{-3.2}$ &
7.2$^{+8.7}_{-5.1}$ (N) & 9.0$^{+4.5}_{-3.2}\times10^{-14}$ &
1.7$^{+0.8}_{-0.6}\times10^{-13}$ & 8.2$_{-2.1}^{+2.8}\times10^{30}$ 
& 1.19 (12)\\
\hline
\end{tabular}
\end{center}
\caption{The spectral parameters derived from fitting an absorbed
multi-temperature thermal plasma model with variable metal abundance,
Z, to the {\xmm} EPIC pn data. The slope of the power law distribution
of temperature, $\alpha$, was fixed at 0.5, and the maximum
temperature, $T_{max}$ of the plasma are shown. We initially set all
the elements to zero and allowed the abundance of each individual
element to vary. We show the observed X-ray flux in the 0.15-10keV
band, $F_{x,o}$, the unabsorbed, bolometric X-ray flux, $F_{x,u}$, the
bolometric X-ray luminosity, $L_{X}$. In determining the luminosities
we assume a distance of 76 pc for V396 Hya and 400 pc for SDSS
J1240-01 (Table \ref{distance}).}
\label{bestfits}
\end{table*}

\section{The X-ray/UV colours of AM CVn systems}

Ramsay et al (2005) showed that AM CVn, HP Lib and CR Boo showed a
much lower soft X-ray to UV ratio compared to most hydrogen accreting
Cataclysmic Variables (CVs).  In contrast, GP Com showed a ratio
similar to that of the hydrogen accreting CVs. We are now able to
increase this sample by including SDSS J1240-01, V396 Hya and ES Cet
(Steeghs et al 2006).

\begin{figure}
\begin{center}
\setlength{\unitlength}{1cm}
\begin{picture}(8,5.)
\put(0.2,-0.2){\includegraphics{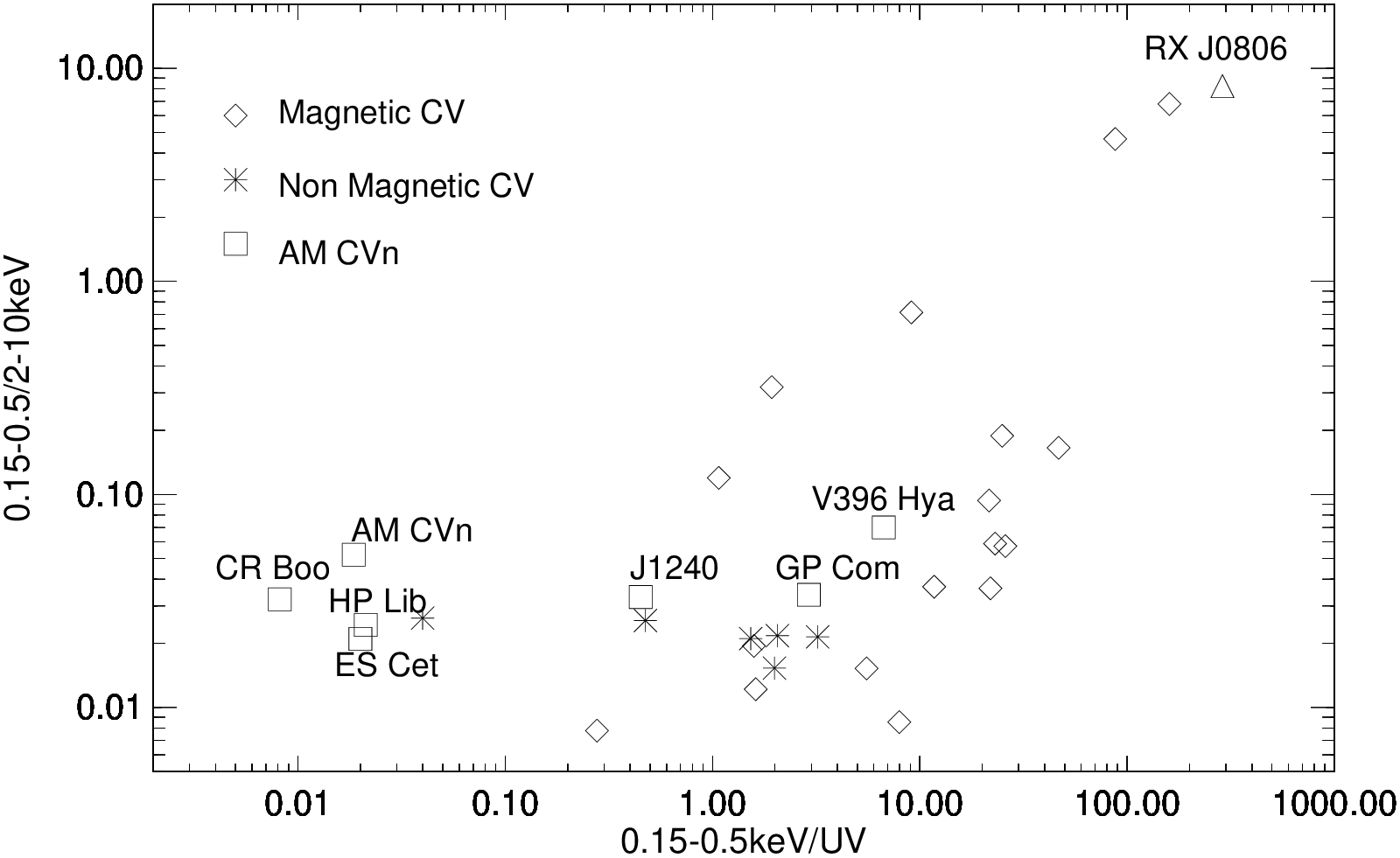}}
\end{picture}
\end{center}
\caption{The colours for the AM CVn systems in this paper;
non-magnetic CVs and the strongly magnetic CVs in the
0.15-0.5/2-10keV, 0.15-0.5/UV plane. We also show the colours of the
candidate system RX J0806+15. The error on the colours are of a
similar size as the symbols.}
\label{colours}
\end{figure}

We adopt the same procedure as Ramsay et al (2005). We determined the
unabsorbed fluxes by setting the absorption parameter in the model
fits discussed earlier to zero and calculated the unabsorbed fluxes in
the 0.15--0.5keV, 2--10keV and UVW1 filter pass bands. These fluxes
were converted to ergs s$^{-1}$ cm$^{-2}$ \AA$^{-1}$ and their colours
in the 0.1--0.5keV/UV, 0.15--0.5/2--10keV plane are shown in Figure
\ref{colours}.

It is noticeable that systems with orbital periods shorter than 25
mins are concentrated in the lower left hand corner of the
colour-colour plane, ie compared to other CVs they show either a low
soft X-ray flux or a high UV flux. In contrast, systems with orbital
periods longer than 25 mins, show a steadily increasing soft X-ray/UV
ratio. We also show the position of the candidate system RX J0806+15,
which is located in the far top right hand corner of the colour-colour
plane. This is not unexpected since there is evidence that this system
could be powered by a non-accretion mechanism and so therefore would
be expected to show different X-ray/UV colours (eg Hakala, Ramsay \&
Byckling 2004). It is predicted that for systems with shorter orbital
periods the emission from an accretion disk will dominate the UV
emission while for longer orbital periods the white dwarf will
dominate (eg Bildsten et al 2006). This would naturally result in the
observed trend in the AM CVn colours seen in Figure \ref{colours}. To
test this we determine the X-ray and UV luminosities in the next
section: the UV luminosity should increase with decreasing orbital
period.

\section{Luminosities}
\label{lum}

Since the paper of Ramsay et al (2005) was published a more accurate
distance has been obtained for AM CVn and distances have been obtained
for the other systems included in this survey. We therefore determine
the X-ray and UV luminosities for our sample using the latest
distances (tabulated in Table \ref{distance}). We note that the quoted
distance to ES Cet was determined using the relationship between
$M_{V}$ and the equivalent width of the He I $\lambda$ 5876
\AA\hspace{1mm} line and therefore is rather uncertain (Espaillat et
al 2005).  The same authors note it could be as high as 1kpc.

The unabsorbed, bolometric, X-ray luminosity was determined using the
model fits to the X-ray spectra shown in Ramsay et al (2005) and in
Table \ref{bestfits}. We assume the X-ray emission is isotropic.
Since we generally have only one UV photometric point from the {\xmm}
OM, determining the UV luminosity is therefore more model dependent
than the X-ray luminosity.  We assume a blackbody of given temperature
then scale the normalisation to match the observed flux at the peak of
the UVW1 filter passband (2910\AA).  Although this method is rather
crude, a comparison between the above method and the flux determined
using the {\sl IUE} spectrum of AM CVn we found that luminosities were
consistent to within 50\%.

A second consideration is correctly modelling the absorption
component.  Since the neutral absorption models in the X-ray spectral
fitting package XSPEC (Arnaud 1996) are not applicable at UV
wavelengths, we also included the {\tt UVRED} model and tied this to
the hydrogen column density parameter using the relationship between
optical extinction and total absorption column of Bohlin, Savage \&
Drake (1978).

We also note that UV emission is likely to be due to a combination of
the heated white dwarf, the accretion stream and accretion disc. While
it is predicted that the disc/stream component will be greater for
shorter periods, taking a mean temperature to represent the UV
emission is clearly somewhat simplistic.  However, observations of CP
Eri showed a temperature of 17000K in the quiescent state (Sion et al
2006). Further, predictions show that the effective temperature is
expected to be in the range of $\sim$10000-40000K (eg Bildsten et al
2006). To determine how sensitive our derived UV luminosities were to
the assumed temperature, we determined the luminosity using blackbody
temperatures between 10000-40000K. We find that compared to a
blackbody of $kT_{bb}$=20000K a blackbody of $kT_{bb}$=10000K gives a
luminosity a factor of 1.3 lower, while a blackbody of
$kT_{bb}$=40000K gives a luminosity a factor of 3.6 greater. We
therefore estimate our UV luminosities will be uncertain by a factor
of $\sim$2-4.

We show the X-ray and UV luminosities in Table \ref{lumin} and in
Figure \ref{lum}. Considering the X-ray luminosity, $L_{X}$, as a
function of orbital period, $P_{orb}$, first. There is no strong
relationship between $L_{X}$ and $P_{orb}$. We do note however, that
AM CVn and HP Lib (with short $P_{orb}$) have higher $L_{X}$ than V396
Hya (the highest $P_{orb}$). The uncertainty to the distance to ES Cet
gives a correspondingly large uncertainty in its $L_{X}$. In contrast,
there is a strong correlation between the UV luminosity, $L_{UV}$, and
$P_{orb}$. $L_{UV}$ increases by 3 orders of magnitude over the
orbital period range of AM CVn systems. This suggests that the
decreasing soft X-ray/UV ratio seen for AM CVn systems in Figure
\ref{colours} is due to the UV luminosity greatly increasing for
shorter binary orbital periods. The distance to RX J0806+15 is rather
uncertain, but the UV luminosity does not follow the relationship for
the other AM CVn systems. Indeed, in contrast to the other sources,
the X-ray luminosity is greater than the UV luminosity. This is
consistent with a non-accretion mechanism powering this source.

\begin{table}
\begin{center}
\begin{tabular}{lrr}
\hline
Source & Distance (pc) & Reference \\
\hline
ES Cet & 350 & Espaillat et al (2005)\\
AM CVn & 606$^{+135}_{-95}$ & Roelofs et al (2006)\\
HP Lib & 197$^{+13}_{-12}$ & Roelofs et al (2006) \\
CR Boo & 337$^{+43}_{-35}$ & Roelofs et al (2006) \\
J1240-01  & 350-440 & Roelofs et al (2005)\\
       & 525 & Bildsten et al (2006)\\
GP Com & 75$\pm2$ & Roelofs et al (2006)\\
V396 Hya & 76$^{+11}_{-8}$ & Thorstensen (Priv Com) \\
\hline
\end{tabular}
\end{center}
\caption{The distances to our sample of AM CVn systems. The distance to 
ES Cet is not certain.}
\label{distance}
\end{table}

\begin{table}
\begin{center}
\begin{tabular}{lrr}
\hline
Source & $L_{UV}$ \ergss & $L_{X}$ \ergss \\
\hline
J0806+15 & $4.5\times10^{31}$  & $1.5\times10^{32}$ \\
ES Cet & $1.4\times10^{34}$  & $8.2\times10^{30}$ \\
AM CVn & $1.1\times10^{34}$  & $2.8\times10^{31}$ \\
HP Lib & $1.0\times10^{34}$  & $1.4\times10^{31}$ \\
CR Boo & $2.7\times10^{33}$  & $5.2\times10^{30}$ \\
J1240-01  & $9.2\times10^{31}$  & $8.2\times10^{30}$ \\
GP Com & $1.5\times10^{31}$  & $5.2\times10^{30}$   \\
V396 Hya & $7.5\times10^{30}$  & $1.7\times10^{30}$ \\
\hline
\end{tabular}
\end{center}
\caption{The X-ray and UV luminosities of our sample of AM CVn
systems. We estimate the UV luminosities are accurate to within a
factor of $\sim$2--4 (see \S {lum}). We assume a distance of 500pc for
RX J0806+15, 1kpc for ES Cet and 400pc for SDSS J1240-01.}
\label{lumin}
\end{table}

\begin{figure}
\begin{center}
\setlength{\unitlength}{1cm}
\begin{picture}(8,5.4)
\put(0,-0.2){\includegraphics{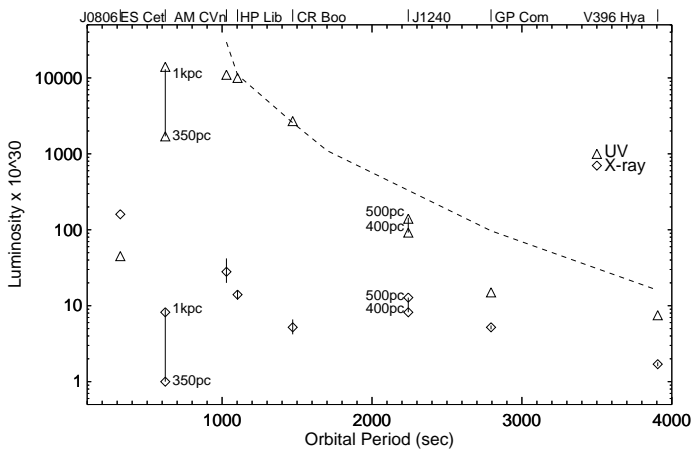}}
\end{picture}
\end{center}
\caption{The (unabsorbed) X-ray and UV luminosities of the AM CVn systems 
included in this sample. The X-ray luminosities were determined using 
modelling the X-ray spectrum derived from {\xmm} data, and the UV luminosities
were determined assuming a blackbody of temperature of 30000K. The 
dotted line is the predicted accretion luminosity -- see \S \ref{discuss_lum}
for details.}
\label{lum}
\end{figure}

\section{Discussion}

\subsection{Abundances}

In the optical, V396 Hya and SDSS J1240-01 appear to show different
abundances, with the former showing helium lines and only a small
number of nitrogen lines (Ruiz et al 2001a), while the latter also
shows silicon and iron lines (Roelofs et al 2005). V396 Hya shows an
optical spectrum similar to GP Com and is thought to be a population
II halo object, while SDSS J1240-01 is thought to be an `ordinary'
population I object, with a higher overall metal abundance (Roelofs et
al 2005).

Our fits to the X-ray spectra of both V396 Hya and SDSS J1240-01 show
a significant enhancement of nitrogen compared to both the solar
abundance and the expected abundance from relatively low temperature
CNO processed material. This is consistent with the fits to the X-ray
spectra of AM CVn, HP Lib, CR Boo and GP Com (Ramsay et al 2005). The
apparent absence of metals other than nitrogen in the X-ray spectrum
of SDSS J1240-01 could be due to the lower signal to noise of the
spectrum compared to the spectrum of V396 Hya.

\subsection{Luminosities and colours}
\label{discuss_lum}

For hydrogen accreting CVs, the orbital evolution is driven by angular
momentum loss via gravitational radiation and magnetic breaking. As
systems evolve to shorter orbital periods the mass transfer rate
decreases by 2-3 orders of magnitude (eg King 1988). In the case of AM
CVn binaries, angular momentum loss is driven entirely by
gravitational radiation. Over the orbital period distribution of AM
CVn systems, this results in a change in the mass transfer rate of
6--7 orders of magnitude, with the highest rate at short orbital
periods (eg Nelemans et al 2001). For both hydrogen CVs and AM CVn
systems, the mass transfer rate is expected to be correlated to the
accretion luminosity.

We find that in AM CVn binaries, the UV luminosity is strongly
correlated with the binary orbital period, with the UV luminosity
increasing as the orbital period gets shorter. The relationship is so
pronounced that it is unlikely to be due to uncertainties in our
method of determining the UV luminosity. This is in contrast to
hydrogen accreting CVs which show the opposite relationship: $L_{UV}$
decreases as the orbital period gets shorter (van Teeseling, Beuermann
\& Verbunt 1996). This is consistent with our view of how these
binaries evolve.

The strong relationship between $L_{UV}$ and $P_{orb}$ for AM CVn
systems allows us to show that the decreasing soft X-ray/UV ratio
shown in Figure \ref{colours} is due to the increasing UV luminosity
of the system. We suggest that this is due to UV emission from the
accretion disk which gets increasingly prominent for systems with
orbital period less than $\sim$25 mins. We note that based on this
relationship, a `high' distance to ES Cet is favoured.

The lack of a strong relationship between the X-ray luminosity and the
orbital period is also noted amongst hydrogen accreting CVs. Only a
weak relationship is seen (van Teeseling, Beuermann \& Verbunt 1996,
Baskill, Wheatley \& Osborne 2005). In both hydrogen accreting systems
and in AM CVn systems, it appears that the change in the accretion
rate is reflected in the UV emission rather than in X-rays.

How does the predicted luminosity compare with the observed X-ray and
UV luminosities? We simplify matters by assuming the accretion
luminosity can be approximated by $L_{acc}=GM_{1}\dot{M}/R$ (in
reality it is expected to be more complex than this) and the mass of
the primary star, $M_{1}$, is taken to be 0.6\Msun (again in reality
$M_{1}$ will span a range in mass). Again, as we noted earlier, some
fraction of the UV flux will originate from the heated white
dwarf. However, at longer orbital periods, where the heated white
dwarf is expected to make a greater contribution to the UV flux, its
luminosity is expected to be several orders of magnitude less than the
solar luminosity, implying that this is a valid assumption (Bildsten
et al 2006).

We have taken the predicted mass transfer rate for different sources
in Table 1 of Deloye, Bildsten \& Nelemans (2005). The mass transfer
rate is set by a number of factors including the mass of the primary
and secondary stars and the secondary stars specific entropy. We took
the mean mass transfer rate for each system and smoothed the resulting
curve of $\dot{M}$ as a function of $P_{orb}$: this is shown in Figure
\ref{lum}. The agreement with the UV luminosity, especially at shorter
orbital periods in remarkable. At longer orbital periods the predicted
luminosity is slightly in excess of the combined X-ray and UV
luminosity, suggesting that for these long period systems, a
significant fraction of the accretion luminosity is emitted at optical
wavelengths.
 
\section{Conclusions}

We have presented X-ray and UV observations of AM CVn binaries which
sample their full orbital period distribution. We find that there is
no evidence for coherent modulations in their X-ray signal. In the UV,
we find strong variability in their signal between orbital periods of
$\sim$1000-1500 sec, although there is no correlation between the
degree of fractional variability and orbital period. 

We find that the soft X-ray/UV ratio decreases as the system orbital
period decreases. This is due to the UV luminosity increasing strongly
as the orbital period decreases. We suggest that this effect is due to
the accretion disc providing an increasingly dominant contribution to
the UV flux as shorter periods. The observed luminosities are in
remarkably good agreement with that predicted and also indicates that
at longer orbital periods a significant fraction of the accretion
energy is also emitted at optical wavelengths.

Our results suggest that new AM CVn systems could be discovered in
X-ray/UV surveys due to their UV variability and their X-ray/UV
colours.

\begin{acknowledgements}

This paper is based on observations obtained using {\sl XMM-Newton},
an ESA science mission with instruments and contributions directly
funded by ESA Member States and the USA (NASA). We thank John
Thorstensen for communicating the distance of V396 Hya prior to
publication. PH is supported by the Academy of Finland, DS
acknowledges a Smithsonian Astrophysical Observatory Clay Fellowship
and support through NASA GO grants NNG04G30G and NNG06GC05G, PJG is
supported by NWO VIDI grant 639.042.201, GN by NWO VENI grant
639.041.405 and TRM was supported by a PPARC Senior Fellowship during
the course of this work.

\end{acknowledgements}

\end{document}